\documentclass[11pt,english,JHEP]{article}
\usepackage{lmodern}
\usepackage[T1]{fontenc}
\usepackage[latin9]{inputenc}
\usepackage{color}
\usepackage{babel}
\usepackage{booktabs}
\usepackage{textcomp}
\usepackage{bm}
\usepackage{amsmath}
\usepackage{amsthm}
\usepackage{amssymb}
\usepackage{graphicx}
\usepackage{setspace}
\usepackage[unicode=true,pdfusetitle,
 bookmarks=true,bookmarksnumbered=false,bookmarksopen=false,
 breaklinks=false,pdfborder={0 0 1},backref=false,colorlinks=true]
 {hyperref}
\hypersetup{
 urlcolor=green, citecolor=blue, linkcolor=blue}

\makeatletter

\providecommand{\tabularnewline}{\\}

\numberwithin{equation}{section}
\theoremstyle{remark}
\newtheorem*{rem*}{\protect\remarkname}
\theoremstyle{plain}
\newtheorem{thm}{\protect\theoremname}
\theoremstyle{definition}
\newtheorem{example}[thm]{\protect\examplename}
\theoremstyle{remark}
\newtheorem*{acknowledgement*}{\protect\acknowledgementname}

\definecolor{green}{RGB}{100, 0, 120}

\newlength{\abstractwidth}
\g@addto@macro\normalsize{%
 \setlength{\abovedisplayskip}{10pt}
 \setlength{\belowdisplayskip}{14pt}
 \setlength{\abovedisplayshortskip}{10pt}
 \setlength{\belowdisplayshortskip}{14pt}}

\usepackage{color}

\makeatother

\providecommand{\acknowledgementname}{Acknowledgement}
\providecommand{\examplename}{Example}
\providecommand{\remarkname}{Remark}
\providecommand{\theoremname}{Theorem}

\begin{document}

\title{\textbf{\huge{}Quantum communication using a quantum switch of quantum
switches }}

\author{Debarshi Das\thanks{S. N. Bose National Centre for Basic Sciences, JD Block, Sector-III,
Bidhannagar, Kolkata 700106, India\textcolor{black}{; }\protect \\
\textcolor{black}{$\hspace{2em}$email: dasdebarshi90@gmail.com}}$\;$\thanks{Department of Physics and Astronomy, University College London, Gower
Street, WC1E 6BT London, UK}$\hspace{1em}\hspace{1em}\hspace{1em}$Somshubhro Bandyopadhyay\thanks{Department of Physics, Bose Institute, Kolkata 700091, India\textcolor{black}{;
email: som.s.bandyopadhyay@gmail.com}}}
\maketitle
\begin{abstract}
The quantum switch describes a quantum operation in which two or more
quantum channels act on a quantum system with the order of application
determined by the state of an order quantum system. And by suitably
choosing the state of the order system, one can create a quantum superposition
of the different orders of application, which can perform communication
tasks impossible within the framework of the standard quantum Shannon
theory. In this paper, we consider the scenario of one-shot heralded
qubit communication and ask whether there exist protocols using a
given quantum switch or switches that could outperform the given ones.
We answer this question in the affirmative. We define a higher-order
quantum switch composed of two quantum switches, with their order
of application controlled by another order quantum system. We then
show that the quantum switches placed in a quantum superposition of
their alternative orders can transmit a qubit, without any error,
with a probability higher than that achievable with the quantum switches
individually. We demonstrate this communication advantage over quantum
switches useful as a resource and those that are useless. We also
show that there are situations where there is no communication advantage
over the individual quantum switches.
\end{abstract}

\section{Introduction }

Quantum theory allows for a novel causal structure with the causal
order between events controlled by a quantum system, which leads to
the quantum superposition of causal orders. The quantum switch \cite{Chiribella+2013},
in particular, combines two or more quantum channels with their order
of application determined by the state of an order quantum system.
For example, in a quantum switch composed of two quantum channels
$\mathcal{E}$ and $\mathcal{F}$, the order is controlled by a qubit.
If the state of this order qubit is $\left|0\right\rangle $, $\mathcal{F}$
is applied before $\mathcal{E},$ but if it is $\left|1\right\rangle $,
then $\mathcal{E}$ is applied before $\mathcal{F}$. However, if
the order qubit is in a superposition of $\left|0\right\rangle $
and $\left|1\right\rangle $, the quantum switch will create a superposition
of the two alternative orders. Then one no longer can determine the
order of application of the channels, and the quantum switch is said
to exhibit \emph{indefinite causal order}, also known as causal nonseparability
\cite{Oreshkov+2012,Araujo+2015,Orsekhkov+2016}. We note that this
superposition is fundamentally different from the type created in
an interferometer that places $\mathcal{E}$ followed by $\mathcal{F}$
on one of the two possible paths and another copy of $\left(\mathcal{E},\mathcal{F}\right)$
with $\mathcal{F}$ followed by $\mathcal{E}$ on the other. The reason
why the superpositions are different lies in the fact that the quantum
switch creates the superposition at the Kraus operator level. Further
note that the quantum switch cannot be realized by a circuit using
quantum channels in a definite causal order or by a statistical mixture
of such circuits \cite{Chiribella 2012}. Recently experimental realizations
of the quantum switch in photonic set-ups have been reported in \cite{exp-procopio+2015,exp-Rubino+2017,exp-Goswami+2018,exp-Guo+2020,exp-Goswami and Romero-2020,exp-Goswami+2020,exp-Rubino+2021}. 

Recent results have established that indefinite causal order is a
bonafide resource for quantum information processing tasks, such as
winning noncausal games \cite{Oreshkov+2012}, testing properties
of quantum channels \cite{Chiribella 2012,Araujo+2014}, reducing
communication complexity \cite{Guerin+2016}, and quantum communication
\cite{Guerin+2016,Ebler+2019,Salek+2018,Chiribella+2021,Bhattacharya+2021,Chribella+2020,Mitra+2021,Abbott+2020}.
The examples demonstrating the communication advantage, however, are
perhaps the most counterintuitive: The quantum switch from two completely
depolarizing channels can transmit classical information \cite{Ebler+2019},
even though the completely depolarizing channel has zero capacity,
and the quantum switch from two completely dephasing channels can
transfer quantum information with nonzero probability although, a
completely dephasing channel cannot \cite{Salek+2018}. Note that
the input qubit in both cases travels through the channels placed
in a superposition of their alternative orders. 

The present paper aims to understand the strengths and limitations
of the quantum switch as a resource for one-shot qubit communication
through noisy channels. Specifically, we investigate what other ways
a quantum switch could be employed to gain an advantage, if any, over
the standard protocol in which a qubit simply traverses the quantum
channels in a superposition of their orders. Now such a strategy in
some cases leads to perfect transfer with a nonzero probability but
in some others, it does not; for example, the quantum switch from
two completely depolarizing channels is useless for the reliable transmission
of a qubit. So for a given quantum channel or collection of channels,
the quantum switch may or may not provide the advantage we seek, and
it is not immediate a priori in which cases it would and in which
it would not. 

One might be tempted to think the standard protocol is perhaps the
only way to obtain a communication advantage, if there is one, using
a given quantum switch. And any other conceivable protocol using the
given switch can be only as good as the standard protocol but not
better. However, we will show that this is not always the case, and
there, in fact, exists a simple protocol using a given switch or a
collection thereof that can outperform the given switch or switches.
We will show this for switches useful as a resource (that is, they
already do better than the quantum channels) and also for those that
are useless. 

We define a higher-order quantum switch, which is both simple and
intuitively satisfying. The definition follows from two observations.
First, the underlying principle of the quantum switch is coherent
control of the order of quantum channels, and second, the quantum
switch itself is a higher-order quantum channel, a bilinear supermap
\cite{Salek+2018,Chiribella+2008,Chiribella+2009}. So one could,
in principle, replace the quantum channels with quantum switches and
thereby construct the higher-order quantum switch, which is what we
do in this paper. This higher-order switch takes two quantum switches
as inputs and creates a new quantum channel, where the order of the
switches is controlled by an order qubit. Then, by choosing the state
of the order qubit in an appropriate manner, one can place the quantum
switches in a quantum superposition of their alternative orders. The
protocols we discuss in this paper exploit this quantum superposition. 

We present three examples where the higher-order quantum switch outperforms
the constituent quantum switches. The first two demonstrate a communication
advantage over useful quantum switches. In particular, we show that
the probability of perfect transfer of an input qubit using the higher-order
quantum switch is greater than that of its constituent quantum switches.
The third one demonstrates a similar advantage over quantum switches
that are useless. We also briefly discuss a possible way to experimentally
realize the higher-order quantum switch using a photonic set-up. 

We wish to point out here that it is not a priori clear that a higher-order
quantum switch, as described above, could indeed provide an advantage
over its constituent quantum switches. The definition seemed natural,
but whether any actual communication advantage could be had was not
guaranteed. In fact, we will show that there are instances in which
we do not observe any communication advantage.

One could consider even higher-order quantum switches. We specifically
discuss the complexity associated with the quantum switches of even
higher-order and the difficulties one would face if one wishes to
employ them for any particular task, not necessarily the one considered
here. 

\section{Quantum switch }

Quantum channels describe the evolution of quantum systems. Mathematically,
a quantum channel is a completely positive, trace-preserving linear
map that transforms quantum states into quantum states. The action
of the channels $\mathcal{E}$ and $\mathcal{F}$ on a quantum state
$\rho$ can be expressed as: 
\begin{alignat}{1}
\mathcal{E}\left(\rho\right) & =\sum_{i}E_{i}\rho E_{i}^{\dagger},\label{E}\\
\mathcal{F}\left(\rho\right) & =\sum_{j}F_{j}\rho F_{j}^{\dagger},\label{F}
\end{alignat}
where $\left\{ E_{i}\right\} $ and $\left\{ F_{j}\right\} $ are
the Kraus operators satisfying $\sum_{i}E_{i}^{\dagger}E_{i}=\sum F_{j}^{\dagger}F_{j}=I$,
$I$ being the identity operator. Suppose the channels are now applied
sequentially. This gives rise to two possible orders: 
\begin{alignat}{1}
\mathcal{F}\circ\mathcal{E}\left(\rho\right) & =\sum_{j,i}F_{j}E_{i}\rho E_{i}^{\dagger}F_{j}^{\dagger},\label{FE}\\
\mathcal{E}\circ\mathcal{F}\left(\rho\right) & =\sum_{i,j}E_{i}F_{j}\rho F_{j}^{\dagger}E_{i}^{\dagger}.\label{EF}
\end{alignat}
Note that, in each of the above scenarios the order in which the channels
are applied to the target state $\rho$ remains fixed. That is, in
(\ref{FE}), $\rho$ is first subjected to $\mathcal{E}$ followed
by $\mathcal{F}$, whereas in (\ref{EF}), it is just the opposite. 

The quantum switch is a higher-order quantum channel constructed from
$\mathcal{E}$, $\mathcal{F}$, and an ancilla $\bm{\omega}$ known
as the order qubit. The order qubit is accessible only to the receiver
and considered to be part of the communication channel. The quantum
switch is defined as \cite{Salek+2018,Chiribella+2008,Chiribella+2009}:
\begin{equation}
\mathbb{S}\left(\mathcal{E},\mathcal{F},\bm{\omega}\right)\left(\rho\right)=\sum_{i,j}K_{ij}\left(\rho\otimes\omega\right)K_{ij}^{\dagger},\label{Switch}
\end{equation}
where $\omega=\left|\omega\right\rangle \left\langle \omega\right|$
is the state of the order qubit and $\left\{ K_{ij}\right\} $ are
the Kraus operators 
\begin{equation}
K_{ij}=E_{i}F_{j}\otimes\left|0\right\rangle \left\langle 0\right|+F_{j}E_{i}\otimes\left|1\right\rangle \left\langle 1\right|,\label{Krauss-Switch}
\end{equation}
where $\left\{ \left|0\right\rangle ,\left|1\right\rangle \right\} $
is an orthonormal basis of $\mathbb{C}^{2}$. Note that (\ref{Switch})
is independent of the Kraus operators $\left\{ E_{i}\right\} $ and
$\left\{ F_{j}\right\} $ \cite{Chiribella+2013}. 

It is evident from (\ref{Krauss-Switch}) that the order in which
the channels $\mathcal{E}$ and $\mathcal{F}$ act is determined by
the state of the order qubit. In particular, if $\left|\omega\right\rangle =\left|0\right\rangle $,
first $\mathcal{F}$ and then $\mathcal{E}$ is applied to $\rho$,
whereas if $\left|\omega\right\rangle =\left|1\right\rangle $ they
are applied in the reverse order. However, if the order qubit is initially
in a superposition state $\left|+\right\rangle =\frac{1}{\sqrt{2}}\left(\left|0\right\rangle +\left|1\right\rangle \right)$
then the quantum switch creates a superposition of the alternative
orders at the Kraus operator level. 

\section{Quantum switch of quantum switches}

Consider the quantum channels $\mathcal{E}^{\left(x\right)}$ with
$\left\{ E_{i}^{\left(x\right)}\right\} $ being the corresponding
set of Kraus operators for $x=1,2,3,4$. Let $\mathbb{S}\left(\mathcal{E}^{\left(1\right)},\mathcal{E}^{\left(2\right)},\bm{\omega}\right)\equiv\mathbb{S}_{1}$
and $\mathbb{S}\left(\mathcal{E}^{\left(3\right)},\mathcal{E}^{\left(4\right)},\bm{\omega}\right)\equiv\mathbb{S}_{2}$
denote the quantum switches from the pairs $\left(\mathcal{E}^{\left(1\right)},\mathcal{E}^{\left(2\right)}\right)$
and $\left(\mathcal{E}^{\left(3\right)},\mathcal{E}^{\left(4\right)}\right)$
respectively. They are defined as 
\begin{align}
\mathbb{S}_{1}\left(\rho\right) & =\sum_{i,j}K_{ij}^{\left(1\right)}\left(\rho\otimes\omega\right)K_{ij}^{\left(1\right)\dagger},\label{Switch-1}\\
\mathbb{S}_{2}\left(\rho\right) & =\sum_{i,j}K_{ij}^{\left(2\right)}\left(\rho\otimes\omega\right)K_{ij}^{\left(2\right)\dagger},\label{Switch-2}
\end{align}
where
\begin{align}
K_{ij}^{\left(1\right)} & =E_{i}^{\left(1\right)}E_{j}^{\left(2\right)}\otimes\left|0\right\rangle \left\langle 0\right|+E_{j}^{\left(2\right)}E_{i}^{\left(1\right)}\otimes\left|1\right\rangle \left\langle 1\right|,\label{Krauss-Switch-1}\\
K_{ij}^{\left(2\right)} & =E_{i}^{\left(3\right)}E_{j}^{\left(4\right)}\otimes\left|0\right\rangle \left\langle 0\right|+E_{j}^{\left(4\right)}E_{i}^{\left(3\right)}\otimes\left|1\right\rangle \left\langle 1\right|.\label{Krauss-Switch-2}
\end{align}

\begin{figure}
\begin{centering}
\includegraphics[scale=0.4]{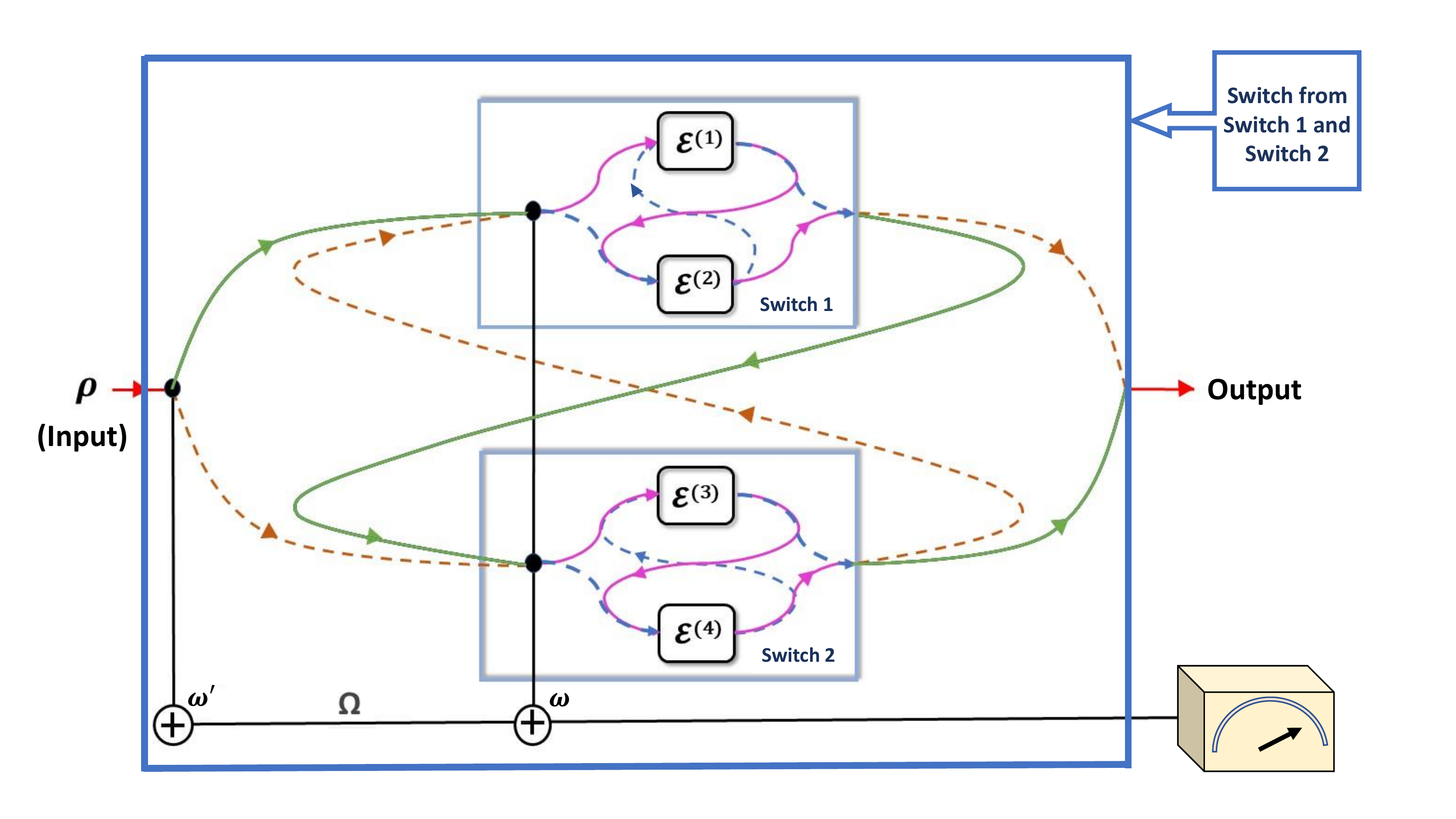}
\par\end{centering}
\textsf{\textcolor{black}{\small{}Figure 1: The mechanism of the higher-order
quantum switch constructed from two quantum switches. Observe that
the order-ancilla consists of two qubits $\bm{\omega}^{\prime}$ and
$\bm{\omega}$. The first qubit controls the order of the switches
whereas the second one controls the order of the channels making up
the individual switches}}\textcolor{black}{\small{}.}{\small\par}
\end{figure}

The quantum switch constructed from the quantum switches $\mathbb{S}_{1}$
and $\mathbb{S}_{2}$ is now defined as: 
\begin{alignat}{1}
\mathbf{S}\left(\mathbb{S}_{1},\mathbb{S}_{2},\bm{\omega}^{\prime}\right)\left(\rho\right) & =\sum_{i,j,k,l}{\rm K}_{ijkl}\left(\rho\otimes\Omega\right)\mathtt{{\rm K}}_{ijkl}^{\dagger},\label{super-switch}
\end{alignat}
where $\bm{\omega}^{\prime}$ is the order qubit controlling the order
of the switches, $\Omega=\left|\Omega\right\rangle \left\langle \Omega\right|$
is the joint state of the two-qubit order ancilla $\bm{\omega\omega}^{\prime}$,
$\bm{\omega}$ being the order qubit associated with $\mathbb{S}_{1}$
and $\mathbb{S}_{2}$, and $\left\{ {\rm K}_{ijkl}\right\} $ are
the Kraus operators defined as
\begin{alignat}{1}
{\rm K}_{ijkl} & =K_{ij}^{\left(1\right)}K_{kl}^{\left(2\right)}\otimes\left|0\right\rangle \left\langle 0\right|+K_{kl}^{\left(2\right)}K_{ij}^{\left(1\right)}\otimes\left|1\right\rangle \left\langle 1\right|.\label{super-Krauss}
\end{alignat}
Observe that the order-ancilla $\bm{\omega\omega}^{\prime}$ is now
a two-qubit system. These two qubits are accessible only to the receiver. 
\begin{rem*}
Since the higher-order switch employs two order qubits, one has the
freedom to choose any initial two-qubit state, which can even be entangled.
Our examples, however, use specific product states, and we did not
find any particular advantage in using entangled states. But we suspect
this may not be the case always, so in situations (like ours or similar)
where the control system is composite, one may examine if an entangled
control has an advantage over a product one.
\begin{rem*}
Our definition of the higher-order quantum switch {[}(\ref{super-switch}){]}
can be viewed as a restricted version of an $N$-switch \cite{Chribella+2020,Mukho_Pati-2020}
for $N=4$, where $N$ is the number of quantum channels. However,
a $4$-switch constructed from the channels $\left\{ \mathcal{E}^{\left(x\right)}\right\} $
comes with a substantial cost in complexity. In particular, allowing
for all possible relative orders, the $4$-switch will lead to a coherent
superposition of all $4!$ alternative orders with the dimension of
the order-system being $4!$. Our definition, on the other hand, is
intuitive, considerably simpler, and physically motivated as a natural
extension of the quantum switch (see Figure 1). 
\end{rem*}
\end{rem*}

\section{Quantum communication using a higher-order quantum switch }

First we will discuss three examples, each of which demonstrates the
relative outperformance of the higher-order quantum switch over constituent
quantum switches. Then we will discuss another example in which the
higher-order quantum switch does not provide any such advantage. 

\subsection{Examples demonstrating communication advantage}
\begin{example}
Consider the Pauli channel $\mathcal{P}$ with the following Kraus
operators:
\begin{equation}
P_{0}=\sqrt{p_{0}}I,\;P_{1}=\sqrt{p_{1}}\sigma_{y},\;P_{2}=\sqrt{p_{2}}\sigma_{z},\label{Pauli-Krauss}
\end{equation}
where $\left(p_{0},p_{1},p_{2}\right)$ is a probability vector with
$0<p_{0},p_{1},p_{2}<1$ and $\sum_{i=0}^{2}p_{i}=1$, and $\sigma_{y}$
and $\sigma_{z}$ are the Pauli $y$ and $z$ matrices respectively.
The action of the Pauli channel on a qubit in state $\rho$ is given
by
\begin{flalign}
\mathcal{P}\left(\rho\right) & =p_{0}\rho+p_{1}\sigma_{y}\rho\sigma_{y}+p_{2}\sigma_{z}\rho\sigma_{z}.\label{Pauli-action}
\end{flalign}
Clearly, for a single use of the Pauli channel error-free qubit communication
is not possible. 

The quantum switch from two Pauli channels $\mathcal{P}$ is defined
as
\begin{equation}
\mathbb{S}\left(\mathcal{P},\mathcal{P},\bm{\omega}\right)\left(\rho\right)=\sum_{i,j=0}^{2}M_{ij}\left(\rho\otimes\omega\right)M_{ij}^{\dagger},\label{Pauli-switch}
\end{equation}
where the Kraus operators $\left\{ M_{ij}\right\} $ are given by
\begin{alignat}{1}
M_{ij} & =P_{i}P_{j}\otimes\left|0\right\rangle \left\langle 0\right|+P_{j}P_{i}\otimes\left|1\right\rangle \left\langle 1\right|,\label{Pauli-switch-Krauss}
\end{alignat}
and $\omega$ is the state of the order qubit $\bm{\omega}$. One
finds that \cite{Chiribella+2021}
\begin{flalign}
\mathbb{S}\left(\mathcal{P},\mathcal{P},\mathfrak{a}\right)\left(\rho\right) & =q_{1}\rho_{1}\otimes\omega+q_{2}\rho_{2}\otimes\sigma_{z}\omega\sigma_{z},\label{Pauli-switch-output}
\end{flalign}
where 
\begin{eqnarray*}
q_{1}=1-2p_{1}p_{2} &  & \rho_{1}=\frac{1}{q_{1}}\left[\left(p_{0}^{2}+p_{1}^{2}+p_{2}^{2}\right)\rho+2p_{0}p_{1}\sigma_{y}\rho\sigma_{y}+2p_{0}p_{2}\sigma_{z}\rho\sigma_{z}\right];\\
q_{2}=2p_{1}p_{2}\hspace{1em}\;\; &  & \rho_{2}=\sigma_{x}\rho\sigma_{x},
\end{eqnarray*}
where $\sigma_{x}$ is the Pauli $x$ matrix. Note that (\ref{Pauli-switch-output})
holds for any $\omega$, so we have the freedom to choose $\omega$
appropriately. Suppose that $\omega=\left|+\right\rangle \left\langle +\right|$;
then $\sigma_{z}\omega\sigma_{z}=\left|-\right\rangle \left\langle -\right|$,
where $\left|-\right\rangle =\frac{1}{\sqrt{2}}\left(\left|0\right\rangle -\left|1\right\rangle \right)$.
From (\ref{Pauli-switch-output}) it follows that if we measure the
order qubit in the $\left\{ \left|\pm\right\rangle \right\} $ basis,
the outcome ``$+$'' will herald the presence of $\rho_{1}$, whereas
the outcome ``$-$'' will herald the presence of $\rho_{2}$. The
first outcome, which occurs with probability $q_{1}$, is not of any
particular interest. But for the second outcome, which occurs with
probability $q_{2}$, an application of $\sigma_{x}$ on the target
qubit leads to an intact transmission of the input state. Thus the
Pauli switch offers a clear advantage over the Pauli channel. Note
that (\ref{Pauli-switch-output}) is valid for any $\omega$. Therefore,
the probability of successful transmission $q_{2}$ cannot be exceeded
for any input state. 

We will now show the higher-order quantum switch constructed from
two Pauli switches can do better. Let us denote the Pauli switch $\mathbb{S}\left(\mathcal{P},\mathcal{P},\bm{\omega}\right)$
by $\mathbb{S}_{\mathcal{P}}$. The higher-order switch is defined
as {[}see (\ref{super-switch}){]}:
\begin{alignat}{1}
\mathbf{S}\left(\mathbb{S}_{\mathcal{P}},\mathbb{S}_{\mathcal{P}},\bm{\omega}^{\prime}\right)\left(\rho\right) & =\sum_{i,j,k,l=0}^{2}{\rm K}_{ijkl}\left(\rho\otimes\Omega\right)\mathtt{{\rm K}}_{ijkl}^{\dagger},\label{Pauli-super-switch}
\end{alignat}
where $\Omega=\left|\Omega\right\rangle \left\langle \Omega\right|$
is the joint state of the order qubits $\bm{\omega}$ and $\bm{\omega}^{\prime}$,
and $\left\{ \text{K}_{ijkl}\right\} $ are the Kraus operators given
by 
\begin{flalign}
\text{K}_{ijkl} & =M_{ij}M_{kl}\otimes\left|0\right\rangle \left\langle 0\right|+M_{kl}M_{ij}\otimes\left|1\right\rangle \left\langle 1\right|.\label{Pauli-super-switch-Krauss}
\end{flalign}
Let us choose $\left|\Omega\right\rangle =\left|++\right\rangle $.
One finds that 
\begin{alignat}{1}
\mathbf{S}\left(\mathbb{S}_{\mathcal{P}},\mathbb{S}_{\mathcal{P}},\bm{\omega}^{\prime}\right)\left(\rho\right) & =\sum_{i=1}^{4}q_{i}\varrho_{i}\otimes\varOmega_{i},\label{output-Pauli-super-switch}
\end{alignat}
where\\
\\
\begin{tabular}{lll}
\toprule 
$\hspace{1em}\hspace{1em}\hspace{1em}\hspace{1em}\hspace{1em}$ $q_{i}$ & $\hspace{1em}\hspace{1em}\hspace{1em}\hspace{1em}\hspace{1em}\hspace{1em}\hspace{1em}\hspace{1em}$
$\varrho_{i}$ & $\hspace{1em}\hspace{1em}\hspace{1em}\hspace{1em}\hspace{1em}$$\varOmega_{i}$\tabularnewline
\midrule
\midrule 
$q_{1}=1-\sum_{i=2}^{4}q_{i}$ & $\varrho_{1}=\left(1-u_{1}-u_{2}\right)\rho+u_{1}\sigma_{y}\rho\sigma_{y}+u_{2}\sigma_{z}\rho\sigma_{z}$ & $\varOmega_{1}=\Omega$\tabularnewline
\midrule 
$q_{2}=8p_{0}^{2}p_{1}p_{2}$ & $\varrho_{2}=\sigma_{x}\rho\sigma_{x}$ & $\varOmega_{2}=\left(I\otimes\sigma_{z}\right)\Omega\left(I\otimes\sigma_{z}\right)$\tabularnewline
\midrule 
$q_{3}=4p_{1}p_{2}\left(p_{0}^{2}+p_{1}^{2}+p_{2}^{2}\right)$ & $\varrho_{3}=\sigma_{x}\rho\sigma_{x}$ & $\varOmega_{3}=\left(\sigma_{z}\otimes I\right)\Omega\left(\sigma_{z}\otimes I\right)$\tabularnewline
\midrule 
$q_{4}=8p_{0}p_{1}p_{2}\left(p_{1}+p_{2}\right)$ & $\varrho_{4}=\left(\frac{p_{2}}{p_{1}+p_{2}}\right)\sigma_{y}\rho\sigma_{y}+\left(\frac{p_{1}}{p_{1}+p_{2}}\right)\sigma_{z}\rho\sigma_{z}$ & $\varOmega_{4}=\left(\sigma_{z}\otimes\sigma_{z}\right)\Omega\left(\sigma_{z}\otimes\sigma_{z}\right)$\tabularnewline
\bottomrule
\end{tabular} \\
\\
and 
\begin{alignat}{1}
u_{1} & =\frac{4p_{0}p_{1}\left(p_{0}^{2}+p_{1}^{2}+p_{2}^{2}\right)}{\left(p_{0}+p_{1}\right)^{4}+4p_{0}p_{2}\left(p_{0}^{2}+p_{1}^{2}+p_{2}^{2}\right)+2p_{2}^{2}\left(3p_{0}^{2}+2p_{0}p_{1}+3p_{1}^{2}\right)+p_{2}^{4}},\label{u1}\\
u_{2} & =\frac{4p_{0}p_{2}\left(p_{0}^{2}+p_{1}^{2}+p_{2}^{2}\right)}{\left(p_{0}+p_{1}\right)^{4}+4p_{0}p_{2}\left(p_{0}^{2}+p_{1}^{2}+p_{2}^{2}\right)+2p_{2}^{2}\left(3p_{0}^{2}+2p_{0}p_{1}+3p_{1}^{2}\right)+p_{2}^{4}}.\label{u2}
\end{alignat}
Now observe that $\varOmega_{2}=\left|+-\right\rangle \left\langle +-\right|$,
$\varOmega_{3}=\left|-+\right\rangle \left\langle -+\right|$, and
$\varOmega_{4}=\left|--\right\rangle \left\langle --\right|$. So
if we measure each order qubit in the $\left\{ \left|\pm\right\rangle \right\} $
basis, the outcomes ``$+-$'' and ``$-+$'' will herald the presence
of $\sigma_{x}\rho\sigma_{x}$. In each case the receiver will be
able to recover the input state simply by applying $\sigma_{x}$ on
the target qubit. This happens with probability
\begin{alignat}{1}
q_{23} & =q_{2}+q_{3}=4p_{1}p_{2}\left(3p_{0}^{2}+p_{1}^{2}+p_{2}^{2}\right).\label{q23}
\end{alignat}
Recall that the Pauli switch achieves noiseless transmission with
probability $2p_{1}p_{2}$. Therefore the higher-order switch will
do better if there exist nonzero $p_{0},p_{1},p_{2}$ satisfying 
\begin{alignat}{1}
4p_{1}p_{2}\left(3p_{0}^{2}+p_{1}^{2}+p_{2}^{2}\right) & >2p_{1}p_{2},\;\sum_{i=0}^{2}p_{i}=1.\label{ineq-1}
\end{alignat}
The above condition is equivalent to 
\begin{flalign}
3p_{0}^{2}+p_{1}^{2}+p_{2}^{2} & >\frac{1}{2},\;\sum_{i=0}^{2}p_{i}=1.\label{ineq-2}
\end{flalign}
It is easy to see that solutions do exist. For example, suppose that
$3p_{0}^{2}=\frac{1}{2}$. Then we need to satisfy 
\begin{alignat}{1}
p_{1}^{2}+p_{2}^{2} & >0,\label{ineq-3}\\
\text{such that\;}p_{1}+p_{2} & =1-\frac{1}{\sqrt{6}},\label{con-3}
\end{alignat}
solutions for which obviously exist. 

\begin{figure}
\begin{centering}
\includegraphics[scale=0.4]{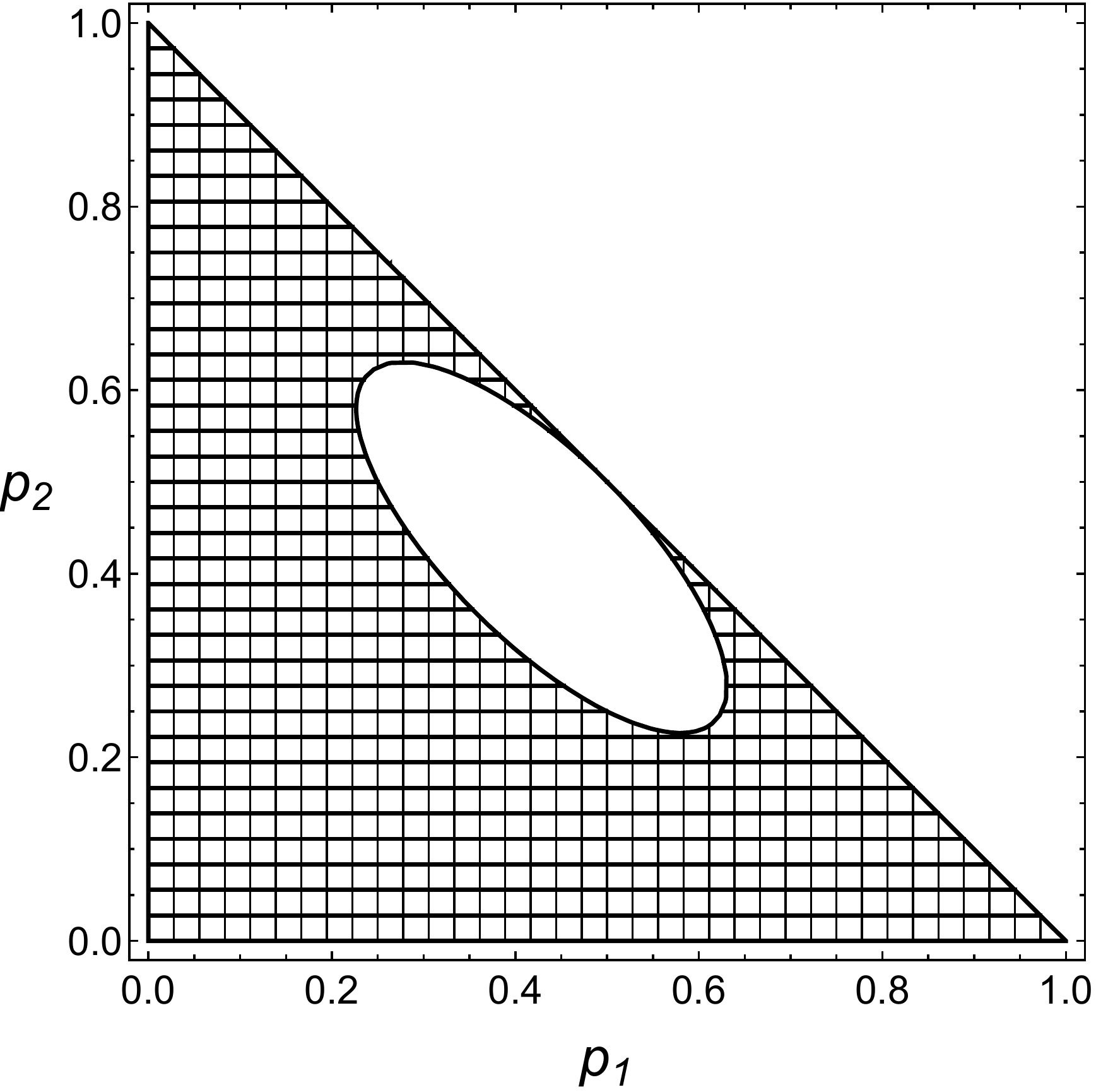}
\par\end{centering}
\textsf{\small{}Figure 2: The shaded region depicts the possible values
of $p_{1}$ and $p_{2}$ for which a single use of the quantum switch
constructed from two Pauli switches enables the noiseless transfer
of an arbitrary qubit with probability higher than the Pauli switch. }{\small\par}
\end{figure}
To find the complete set of solutions, define 
\begin{flalign}
\Delta & =q_{23}-2p_{1}p_{2}\nonumber \\
 & =2p_{1}p_{2}\left[2\left\{ 3\left(1-p_{1}-p_{2}\right)^{2}+p_{1}^{2}+p_{2}^{2}\right\} -1\right],\label{delta}
\end{flalign}
where to arrive at the last line we have used $p_{0}=1-p_{1}-p_{2}$.
Then all possible pairs $\left(p_{1},p_{2}\right)$, where $0<p_{1},p_{2}<1$
and $p_{1}+p_{2}<1$, for which $\Delta>0$ are admissible solutions
(see, Figure 2). 
\begin{example}
Consider the bit flip channel $\mathcal{B}$ with the Kraus operators
\begin{equation}
B_{0}=\sqrt{1-r}I,\hspace{1em}B_{1}=\sqrt{r}\sigma_{x},\;0<r<1,\label{Bit-Krauss}
\end{equation}
and the phase flip channel $\mathcal{G}$ with the Kraus operators
\begin{equation}
G_{0}=\sqrt{1-s}I,\hspace{1em}G_{1}=\sqrt{s}\sigma_{z},\;0<s<1.\label{Phase-Krauss}
\end{equation}
Their actions on a qubit state $\rho$ are written as
\begin{flalign}
\mathcal{B}\left(\rho\right) & =\left(1-r\right)\rho+r\sigma_{x}\rho\sigma_{x},\label{Bit-action}\\
\mathcal{G}\left(\rho\right) & =\left(1-s\right)\rho+s\sigma_{z}\rho\sigma_{z}.\label{Phase-action}
\end{flalign}
It is clear that for single use of the above channels error-free transfer
of a qubit state is not possible. 

Consider now the quantum switch constructed from $\mathcal{B}$ and
$\mathcal{G}$:
\begin{flalign}
\mathbb{S}\left(\mathcal{B},\mathcal{G},\bm{\omega}\right)\left(\rho\right) & =\sum_{i,j=0}^{1}T_{ij}\left(\rho\otimes\omega\right)T_{ij}^{\dagger},\label{BG-switch}
\end{flalign}
where 
\begin{alignat}{1}
T_{ij} & =B_{i}G_{j}\otimes\left|0\right\rangle \left\langle 0\right|+G_{j}B_{i}\otimes\left|1\right\rangle \left\langle 1\right|\label{BG=00003Dswitch-Krauss}
\end{alignat}
are the Kraus operators. Simplifying (\ref{BG-switch}) one obtains
\begin{alignat}{1}
\mathbb{S}\left(\mathcal{B},\mathcal{G},\bm{\omega}\right)\left(\rho\right) & =\left[\left(1-r\right)\left(1-s\right)\rho+r\left(1-s\right)\sigma_{x}\rho\sigma_{x}+s\left(1-r\right)\sigma_{z}\rho\sigma_{z}\right]\otimes\omega\nonumber \\
 & \;\;\;+rs\left(\sigma_{y}\rho\sigma_{y}\right)\otimes\left(\sigma_{z}\omega\sigma_{z}\right).\label{eq:BG-switch-output}
\end{alignat}
For $\omega=\left|+\right\rangle \left\langle +\right|$ we have $\sigma_{z}\omega\sigma_{z}=\left|-\right\rangle \left\langle -\right|$.
From (\ref{eq:BG-switch-output}) it follows that measuring the order
qubit in the $\left\{ \left|\pm\right\rangle \right\} $ basis, the
outcome ``$-$'' will be obtained with probability $rs$. For this
outcome, the target qubit ends up in the state $\sigma_{y}\rho\sigma_{y}$
and can be subsequently recovered by applying $\sigma_{y}$. Thus
perfect transfer of an arbitrary qubit is possible with probability
$rs$ using $\mathbb{S}\left(\mathcal{B},\mathcal{G},\bm{\omega}\right)$
only once. Since (\ref{eq:BG-switch-output}) holds for any initial
state $\omega$, the probability of noiseless transmission cannot
be exceeded. 

Let us now consider the higher-order quantum switch composed of two
identical quantum switches $\mathbb{S}\left(\mathcal{B},\mathcal{G},\bm{\omega}\right)$
denoted by $\mathbb{S}_{\mathcal{B},\mathcal{G}}$. The higher-order
switch is defined as {[}(\ref{super-switch}){]}: 
\begin{alignat}{1}
\mathbf{S}\left(\mathbb{S}_{\mathcal{B},\mathcal{G}},\mathbb{S}_{\mathcal{B},\mathcal{G}},\bm{\omega}^{\prime}\right)\left(\rho\right) & =\sum_{i,j,k,l=0}^{1}{\rm K}_{ijkl}\left(\rho\otimes\Omega\right)\mathtt{{\rm K}}_{ijkl}^{\dagger},\label{BG-super-switch}
\end{alignat}
where $\Omega=\left|\Omega\right\rangle \left\langle \Omega\right|$
is the joint state of the order qubits $\bm{\omega}$ and $\bm{\omega}^{\prime}$,
and $\left\{ \text{K}_{ijkl}\right\} $ are the Kraus operators given
by 
\begin{flalign}
\text{K}_{ijkl} & =T_{ij}T_{kl}\otimes\left|0\right\rangle \left\langle 0\right|+T_{kl}T_{ij}\otimes\left|1\right\rangle \left\langle 1\right|.\label{BG-super-switch-Krauss}
\end{flalign}
Let us now choose $\left|\Omega\right\rangle =\left|++\right\rangle $.
One obtains 
\begin{alignat}{1}
\mathbf{S}\left(\mathbb{S}_{\mathcal{B},\mathcal{G}},\mathbb{S}_{\mathcal{B},\mathcal{G}},\bm{\omega}^{\prime}\right)\left(\rho\right) & =\sum_{i=1}^{4}q_{i}^{\prime}\varrho_{i}^{\prime}\otimes\varOmega_{i}^{\prime},\label{BG-super-switch-action}
\end{alignat}
where\\
\\
\begin{tabular}{lll}
\toprule 
$\hspace{1em}\hspace{1em}\hspace{1em}\hspace{1em}\hspace{1em}$ $q_{i}^{\prime}$  & $\hspace{1em}\hspace{1em}\hspace{1em}\hspace{1em}\hspace{1em}\hspace{1em}\hspace{1em}\hspace{1em}$
$\varrho_{i}^{\prime}$ & $\hspace{1em}\hspace{1em}\hspace{1em}\hspace{1em}\hspace{1em}$$\varOmega_{i}^{\prime}$\tabularnewline
\midrule
\midrule 
$q_{1}^{\prime}=1-\sum_{i=2}^{4}q_{i}^{\prime}$ & $\varrho_{1}^{\prime}=\left(1-v_{1}-v_{2}\right)\rho+v_{1}\sigma_{x}\rho\sigma_{x}+v_{2}\sigma_{z}\rho\sigma_{z}$ & $\varOmega_{1}^{\prime}=\Omega$\tabularnewline
\midrule 
$q_{2}^{\prime}=2rs\left(1-r\right)\left(1-s\right)$ & $\varrho_{2}^{\prime}=\sigma_{y}\rho\sigma_{y}$ & $\varOmega_{2}^{\prime}=\left(I\otimes\sigma_{z}\right)\Omega\left(I\otimes\sigma_{z}\right)$\tabularnewline
\midrule 
$q_{3}^{\prime}=q_{2}^{\prime}$ & $\varrho_{3}^{\prime}=\sigma_{y}\rho\sigma_{y}$  & $\varOmega_{3}^{\prime}=\left(\sigma_{z}\otimes I\right)\Omega\left(\sigma_{z}\otimes I\right)$\tabularnewline
\midrule 
$q_{4}^{\prime}=2rs\left(r+s-2rs\right)$ & $\varrho_{4}^{\prime}=\frac{1}{\left(r+s-2rs\right)}\left[s\left(1-r\right)\sigma_{x}\rho\sigma_{x}+r\left(1-s\right)\sigma_{z}\rho\sigma_{z}\right]$ & $\varOmega_{4}^{\prime}=\left(\sigma_{z}\otimes\sigma_{z}\right)\Omega\left(\sigma_{z}\otimes\sigma_{z}\right)$\tabularnewline
\bottomrule
\end{tabular}\\
\\
and
\begin{alignat}{1}
v_{1} & =\frac{2r\left(1-r\right)\left(1-s\right)^{2}}{1-2rs\left(2-r-s\right)},\label{v1}\\
v_{2} & =\frac{2s\left(1-s\right)\left(1-r\right)^{2}}{1-2rs\left(2-r-s\right)}.\label{v2}
\end{alignat}
Observe that $\varOmega_{2}^{\prime}=\left|+-\right\rangle \left\langle +-\right|$,
$\varOmega_{3}^{\prime}=\left|-+\right\rangle \left\langle -+\right|$,
and $\varOmega_{4}^{\prime}=\left|--\right\rangle \left\langle --\right|$.
Therefore, measuring the order qubits in the $\left\{ \left|\pm\right\rangle \right\} $
basis will herald the presence of $\sigma_{y}\rho\sigma_{y}$ for
the outcomes ``$+-$'' and ``$-+$'', and the receiver can now
recover the input state by applying $\sigma_{y}$. Hence, with the
probability
\begin{alignat}{1}
q_{23}^{\prime} & =q_{2}^{\prime}+q_{3}^{\prime}=4rs\left(1-r\right)\left(1-s\right)\label{qprime-23}
\end{alignat}
the higher-order switch achieves error-free transmission of an input
qubit. 

\begin{figure}
\begin{centering}
\includegraphics[scale=0.4]{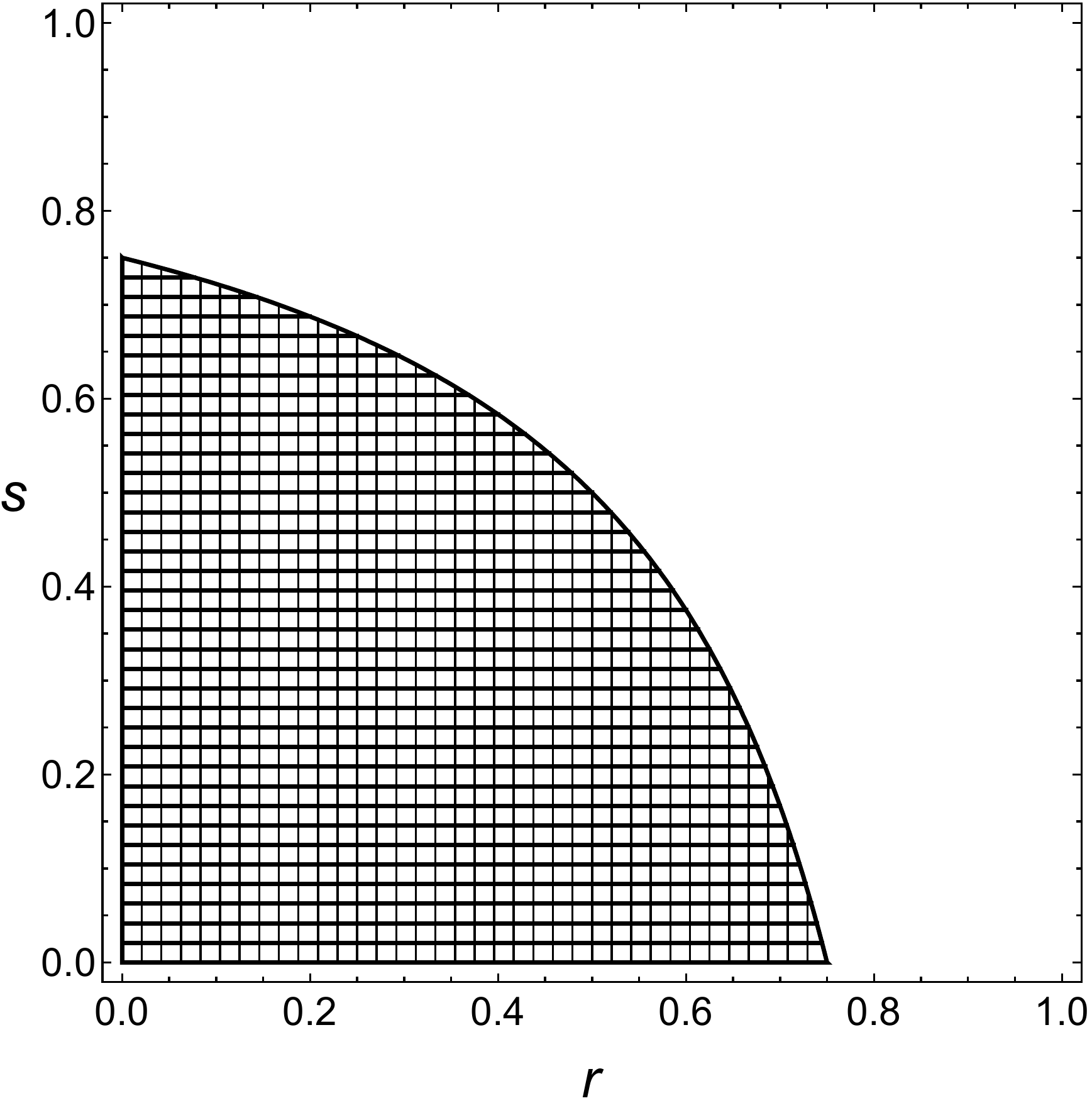}
\par\end{centering}
\textsf{\small{}Figure 3: The shaded region depicts the possible values
of $r$ and $s$ for which a single use of the quantum switch from
two quantum switches $\mathbb{S}\left(\mathcal{B},\mathcal{G},\mathfrak{a}\right)$
enables the noiseless transfer of an arbitrary qubit with probability
higher than $\mathbb{S}\left(\mathcal{B},\mathcal{G},\mathfrak{a}\right)$.}\texttt{\small{} }{\small\par}
\end{figure}
 Th higher-order switch will outperform the constituent switches provided
we can find $0<r,s<1$ such that $q_{23}^{\prime}>rs$. That means
we need to satisfy the inequality 
\begin{alignat}{1}
\left(1-r\right)\left(1-s\right) & >\frac{1}{4}.\label{r-s-ineq-2}
\end{alignat}
One can easily see that solutions do exist. For example, for all $0<r,s<\frac{1}{2}$,
(\ref{r-s-ineq-2}) will be satisfied. The complete set of admissible
solutions is depicted in Figure 3. 
\begin{example}
Consider now two quantum switches, the first constructed from two
identical bit flip channels $\mathcal{B}$ and the second from two
identical phase flip channels $\mathcal{G}$:

\begin{flalign}
\mathbb{S}\left(\mathcal{B},\mathcal{B},\bm{\omega}\right)\left(\rho\right) & =\sum_{i,j=0}^{1}V_{ij}\left(\rho\otimes\omega\right)V_{ij}^{\dagger},\label{BB-switch}\\
\mathbb{S}\left(\mathcal{G},\mathcal{G},\bm{\omega}\right)\left(\rho\right) & =\sum_{i,j=0}^{1}W_{ij}\left(\rho\otimes\omega\right)W_{ij}^{\dagger},\label{GG-switch}
\end{flalign}
where 
\begin{alignat}{1}
V_{ij} & =B_{i}B_{j}\otimes\left|0\right\rangle \left\langle 0\right|+B_{j}B_{i}\otimes\left|1\right\rangle \left\langle 1\right|\label{BB-switch-Krauss}\\
W_{ij} & =G_{i}G_{j}\otimes\left|0\right\rangle \left\langle 0\right|+G_{j}G_{i}\otimes\left|1\right\rangle \left\langle 1\right|\label{GG-switch-Krauss}
\end{alignat}
are the corresponding Kraus operators. 

Simplifying (\ref{BB-switch-Krauss}) and (\ref{GG-switch-Krauss})
one finds that 
\begin{flalign}
\mathbb{S}\left(\mathcal{B},\mathcal{B},\bm{\omega}\right)\left(\rho\right) & =\left[\left(1-b\right)\rho+b\sigma_{x}\rho\sigma_{x}\right]\otimes\omega,\label{BB-switch-1}\\
\mathbb{S}\left(\mathcal{G},\mathcal{G},\bm{\omega}\right)\left(\rho\right) & =\left[\left(1-g\right)\rho+g\sigma_{z}\rho\sigma_{z}\right]\otimes\omega,\label{GG-switch-1}
\end{flalign}
where $b=2r\left(1-r\right)$ and $g=2s\left(1-s\right)$. Since (\ref{BB-switch-1})
and (\ref{GG-switch-1}) hold for an arbitrary $\omega$, we conclude
that neither switch can transfer a qubit without error when used only
once. 

Let us now consider the higher-order quantum switch composed of $\mathbb{S}\left(\mathcal{B},\mathcal{B},\bm{\omega}\right)$
and $\mathbb{S}\left(\mathcal{G},\mathcal{G},\bm{\omega}\right)$.
Let us denote the constituent switches by $\mathbb{S}_{\mathcal{B}}$
and $\mathbb{S}_{\mathcal{G}}$ respectively. The higher-order switch
is defined as:
\begin{alignat}{1}
\mathbf{S}\left(\mathbb{S}_{\mathcal{B}},\mathbb{S}_{\mathcal{G}},\bm{\omega}^{\prime}\right)\left(\rho\right) & =\sum_{i,j,k,l=0}^{2}{\rm K}_{ijkl}\left(\rho\otimes\Omega\right)\mathtt{{\rm K}}_{ijkl}^{\dagger},\label{BG-super-switch-1}
\end{alignat}
where $\Omega=\left|\Omega\right\rangle \left\langle \Omega\right|$
is the joint state of the order qubits $\bm{\omega}$ and $\bm{\omega}^{\prime}$,
and $\left\{ \text{K}_{ijkl}\right\} $ are the Kraus operators given
by 
\begin{flalign}
\text{K}_{ijkl} & =V_{ij}W_{kl}\otimes\left|0\right\rangle \left\langle 0\right|+W_{kl}V_{ij}\otimes\left|1\right\rangle \left\langle 1\right|.\label{BG-super-switch-Krauss-1}
\end{flalign}
Let us now choose $\left|\Omega\right\rangle =\left|++\right\rangle $.
Then, we get 
\begin{alignat}{1}
\mathbf{S}\left(\mathbb{S}_{\mathcal{B}},\mathbb{S}_{\mathcal{G}},\bm{\omega}^{\prime}\right)\left(\rho\right) & =\sum_{i=1}^{2}q_{i}^{\prime\prime}\varrho_{i}^{\prime\prime}\otimes\varOmega_{i}^{\prime\prime},\label{BG-super-switch-action-1}
\end{alignat}
where\\
\\
\begin{tabular}{lll}
\toprule 
$\hspace{1em}\hspace{1em}\hspace{1em}\hspace{1em}\hspace{1em}$ $q_{i}^{\prime\prime}$  & $\hspace{1em}\hspace{1em}\hspace{1em}\hspace{1em}\hspace{1em}\hspace{1em}\hspace{1em}\hspace{1em}$
$\varrho_{i}^{\prime\prime}$ & $\hspace{1em}\hspace{1em}\hspace{1em}\hspace{1em}\hspace{1em}$$\varOmega_{i}^{\prime\prime}$\tabularnewline
\midrule
\midrule 
$q_{1}^{\prime\prime}=1-q_{2}^{\prime\prime}$ & $\varrho_{1}^{\prime\prime}=\left(1-w_{1}-w_{2}\right)\rho+w_{1}\sigma_{x}\rho\sigma_{x}+w_{2}\sigma_{z}\rho\sigma_{z}$ & $\varOmega_{1}^{\prime\prime}=\Omega$\tabularnewline
\midrule 
$q_{2}^{\prime\prime}=4rs\left(1-r\right)\left(1-s\right)$ & $\varrho_{2}^{\prime\prime}=\sigma_{y}\rho\sigma_{y}$ & $\varOmega_{2}^{\prime\prime}=\left(I\otimes\sigma_{z}\right)\Omega\left(I\otimes\sigma_{z}\right)$\tabularnewline
\bottomrule
\end{tabular}\\
\\
and
\begin{alignat}{1}
w_{1} & =\frac{2\left[2\left(1-s\right)s-1\right]\left(1-r\right)r}{4rs\left(1-r\right)\left(1-s\right)-1},\label{v1-1}\\
w_{2} & =\frac{2\left[2\left(1-r\right)r-1\right]\left(1-s\right)s}{4rs\left(1-r\right)\left(1-s\right)-1}.\label{v2-1}
\end{alignat}
Since $\Omega_{2}^{\prime\prime}=\left|+-\right\rangle \left\langle +-\right|$,
measuring each of the two order qubits in the $\left\{ \left|\pm\right\rangle \right\} $
basis will herald the presence of $\sigma_{y}\rho\sigma_{y}$ whenever
the outcome is ``$+-$''. This happens with probability $q_{2}^{\prime\prime}=4rs\left(1-r\right)\left(1-s\right)$
and when it does, the input state can be recovered by applying $\sigma_{y}$.
So with probability $q_{2}^{\prime\prime}$ one achieves error-free
transfer of a qubit for single use of the higher-order switch. 

Recall that the switches $\mathbb{S}\left(\mathcal{B},\mathcal{B},\mathfrak{a}\right)$
and $\mathbb{S}\left(\mathcal{G},\mathcal{G},\mathfrak{a}\right)$
are completely useless, for they cannot achieve perfect transmission
of a qubit with a nonzero probability. But, as we have just shown,
the higher-order switch can perform this task with a nonzero probability
for all $0<r,s<1$ . So a higher-order quantum switch, even if composed
of useless quantum switches, can function as a resource. 
\end{example}

\end{example}

\end{example}

\subsection{No communication advantage using a higher-order switch}

The higher-order quantum switch, however, cannot always outperform
the quantum switches from which it has been constructed. Consider
the bit flip channel $\mathcal{B}^{1/2}$ and the phase flip channel
$\mathcal{G}^{1/2}$ defined by (\ref{Bit-Krauss}) and (\ref{Phase-Krauss})
for $r=1/2$ and $s=1/2$ , respectively. While the channels are useless
for noiseless transfer of a qubit, the quantum switch $\mathbb{S}\left(\mathcal{B}^{1/2},\mathcal{G}^{1/2},\bm{\omega}\right)$
constructed from $\mathcal{B}^{1/2}$ and $\mathcal{G}^{1/2}$ is
useful as a resource. In particular, it allows for noiseless transfer
of a qubit in the single-shot case with probability $1/4$ \cite{Salek+2018}.
However, the higher-order switch $\mathbf{S}$ composed of two identical
switches $\mathbb{S}\left(\mathcal{B}^{1/2},\mathcal{G}^{1/2},\bm{\omega}\right)$
fails to do any better. One can show that the probability for noiseless
transfer of an input qubit using the higher-order switch does not
exceed $1/4$ for any choice of the joint state $\left|\Omega\right\rangle $
of the order qubits. The calculation is similar to the above examples,
so we do not give it here. 

\section{Discussions}

In this section, we briefly discuss two things. First, the complexity
associated with the quantum switches of even higher-orders and next,
a possible way to experimentally implement the higher-order quantum
switch.

\subsection{Quantum switches of even higher-orders }

The definition of the higher-order quantum switch presented here can
easily be generalized to construct even higher-order quantum switches.
Clearly, the dimension of the order quantum system will grow, and
so will the number of Kraus operators. For example, consider a quantum
channel with $m\geq2$ Kraus operators. The quantum switch of two
copies of the quantum channel will therefore have $\left(m\times m\right)$
Kraus operators. Let us call this the first-order quantum switch $\mathbb{S}_{1}$.
Then the second order quantum switch $\mathbb{S}_{2}$ constructed
from two copies of $\mathbb{S}_{1}$ will have $\left(m\times m\right)\times\left(m\times m\right)=\left(m\times m\right)^{2}$
Kraus operators. The definition and examples presented in this study
are nothing but this kind of second order quantum switch. Generalizing
this approach, the $n^{\text{th}}$ order quantum switch $\mathbb{S}_{n}$
constructed from two identical $\left(n-1\right)^{\text{th}}$ order
switches $\mathbb{S}_{n-1}$ will have $\left(m\times m\right)^{2^{n-1}}$
Kraus operators. For example, if one simply moves one step up from
the situations we have considered in the Examples 2 and 3, the resulting
higher-order quantum switch will require 256 Kraus operators. 

Besides, the dimension of the order-ancilla will grow as well. And
that leads to a problem of a different kind. Following our definition,
the state of the order-ancilla associated with the $\mathbb{S}_{n}$
will be an $n$-qubit state (where the same order qubit is assumed
to be associated with both copies of $\mathbb{S}_{n-1}$ making up
$\mathbb{S}_{n}$). That means, to analyze such a situation with full
generality, one must consider all possible $n$ qubit initial states. 

So we see that there is a hierarchy and as we move up the complexity
grows pretty quickly. Such higher-order quantum switches could indeed
provide the communication advantage over the lower order ones, but
it is not at all clear in which cases it would and in which it would
not. Furthermore, we suspect the communication advantage, if any,
could be incremental or even capped, but none of these suppositions
are clear at this point.

\subsection{Experimental realization of the higher-order quantum switch}

The results presented here can be implemented by suitably modifying
and generalizing the techniques adopted in the recent experiments
\cite{exp-procopio+2015,exp-Rubino+2017,exp-Goswami+2018,exp-Guo+2020,exp-Goswami and Romero-2020,exp-Goswami+2020,exp-Rubino+2021}.
For example, following \cite{exp-Rubino+2021}, the input qubit can
be realized through the internal polarization degree of freedom associated
with a photon. The effects of the different types of channels considered
in this paper can then be implemented using liquid-crystal wave plates
that can rapidly implement different polarization rotations \cite{exp-Rubino+2021}.
Each quantum switch can be created through a Mach-Zehnder interferometer
that creates a quantum superposition of the path degrees of freedom
using a $50:50$ beam splitter. Here the order-qubit is realized via
the path degrees of freedom. In particular, the states $\left|0\right\rangle $
and $\left|1\right\rangle $ of the order qubit correspond to a photon
travelling the first arm and the second arm of the interferometer,
respectively. Then to realize a quantum switch the interferometer
needs to be folded into two loops, so the photon can now travel through
the two channels in the two alternative orders in each arm of the
interferometer (see, \cite{exp-Rubino+2021} for details). Note that
in such a folded interferometer, only one copy of each of the two
channels is used in such a way that the photon can travel through
the two channels in two alternative orders by travelling through the
two arms. That is, here the same copy of any of the two channels acts
on the photon when it travels through the first arm or through the
second arm. This cannot be achieved simply by using an interferometer.
Rather, the aforementioned ``folded'' interferometer is required
to realize this.

Therefore, in order to create quantum switches $\mathbb{S}_{1}$ and
$\mathbb{S}_{2}$, two folded Mach-Zehnder interferometers $\text{I}_{1}$
and $\text{I}_{2}$ are necessary. The quantum switch of two quantum
switches can now be implemented using a third folded Mach-Zehnder
interferometer $\text{I}_{3}$. Here, the two quantum switches (the
Mach-Zehnder interferometers $\text{I}_{1}$ and $\text{I}_{2}$)
need to appear in alternating order in each of the arms of the interferometer
$\text{I}_{3}$. To summarize, the input qubit can be encoded in the
polarization degrees of freedom, and the order-ancilla can be realized
via the path degrees of freedom. Finally, each of the order qubits
can be measured in the Hadamard basis by suitably setting relative
phases between the different arms before recombining them at the end
of the interferometers.

\section{Conclusions}

The quantum switch leads to a novel causal structure where two quantum
channels act in a quantum superposition of their possible causal orders
\cite{Chiribella+2013}. It has also been shown that the indefinite
causal order manifested in a quantum switch is a resource for quantum
communication \cite{Ebler+2019,Salek+2018,Chiribella+2021,Bhattacharya+2021,Chribella+2020}.
In this paper, we discussed a higher-order quantum switch. Here, a
quantum state could pass through two quantum switches in a superposition
of different causal orders, where the order of the quantum switches
is controlled by an order qubit. We showed that in one-shot heralded
quantum communication this higher-order quantum switch can perform
better than the individual quantum switches. In particular, two quantum
switches placed in a quantum superposition of their alternative causal
orders can transmit a qubit without any error with a probability higher
than that of the individual quantum switches. We discussed three examples
in detail. The first two showed this outperformance over useful quantum
switches whereas, the last one showed that a higher-order quantum
switch becomes useful even when constructed from two useless quantum
switches. However, this outperformance is not something that is given.
We have discussed a specific example where the higher-order quantum
switch does not outperform the constituent quantum switches. We also
discussed a way to realize the higher-order quantum switch in an experiment
and the complexity of even higher-order quantum switches. 

There, however, are situations where the communication advantage using
a quantum switch (where the order of the channels is coherently controlled)
can also be obtained using coherently controlled quantum channels
(where the choice of the quantum channel is coherently controlled)
without requiring indefinite causal order \cite{Guerin+2019,Abbott+2020},
although this is not always the case \cite{Bhattacharya+2021}. So
it would be interesting to find out whether our results or similar
ones can also be reproduced in a set-up of coherently controlled quantum
channels without involving indefinite causal order. We do not know
the answer either way and leave it for future consideration. 

There is another problem one might consider. It has been shown there
exist noisy quantum channels that can act as a perfect quantum communication
channel when used to form a quantum switch \cite{Chiribella+2021}.
However, these examples are unique up to unitary freedom. In a practical
situation, it could be the case that particular quantum channels are
not available. This stipulates the question: How to improve the efficacy
of a quantum switch from two arbitrary noisy quantum channels? We
answered this question partially by showing that the performances
of certain quantum switches can be improved by a higher-order quantum
switch. But there could be other ways to achieve the same, so other
ideas also need to be explored. So, it would be interesting to know
whether a higher-order quantum switch from two \textquotedblleft noisy\textquotedblright{}
quantum switches can behave as a perfect quantum communication channel. 
\begin{acknowledgement*}
DD acknowledges Ananda G. Maity for fruitful discussion and the Science
and Engineering Research Board (SERB), Government of India for financial
support through a National Post Doctoral Fellowship (File No.: PDF/2020/001358).
During the later phase of this work, the research of DD is supported
by the Royal Society (United Kingdom) through the Newton International
Fellowship (NIF/R1/212007).
\end{acknowledgement*}

\end{document}